\documentclass[11pt,a4paper]{article}
\usepackage{amsmath}

\usepackage[psamsfonts]{amssymb}
\usepackage{amsmath}
\usepackage{epsfig}
\usepackage{float}

\author{H. M. Sadjadi\footnote{M.sad@ut.ac.ir}
\\ {\small Department of Physics, University of Tehran,}
\\ {\small P. O. B. 14395-547, Tehran 14399-55961, Iran}}
\author{H. M. Sadjadi \footnote{M.sad@ut.ac.ir}
\\ {\small Department of Physics, University of Tehran,}
\\ {\small P. O. B. 14395-547, Tehran 14399-55961, Iran}}
\title{On solutions of loop quantum cosmology}
\begin{document}
\maketitle
\begin{abstract}
Loop quantum cosmology is considered in inflation era. A slow roll
scalar field solution with power law potential is presented in the
neighborhood of transition time, i.e. when the universe enters
inflation phase from super-inflation. The compatibility of the
model with Planck 2013 data is discussed. The domain of validity
of second and the generalized second laws of thermodynamics for
this solution and some other examples is studied.
\end{abstract}

\section{Introduction}

Application of loop quantum gravity (LQG) to homogenous and
isotropic space time, reduces its symmetries and gives rise to
loop quantum cosmology (LQC) \cite{lqc}.  In this theory, due to
the effective quantum gravitational effects, the big bang
singularity predicted by general relativity is replaced by a big
bounce, after which a super-inflation phase occurs. Afterwards the
universe enters a normal inflation regime.

By super-inflation we mean a period of super-acceleration of the
universe during which the time derivative of the Hubble parameter,
$H$,  is positive: $\dot{H}>0$, in contrast to the standard slow
roll inflation where $\dot{H}<0$. This super-acceleration can be
viewed as a special feature of modified Friedmann equations in
effective theory of loop quantum cosmology. Such a phase can also
be occurred in general theory of gravity but for example by
introducing exotic matter such phantom which violates the null
energy condition, or by adding non-minimal coupling terms to the
matter action. By considering the super-inflation, the horizon
problem may be solved with only a few e-folds \cite{sha}.
Perturbations in this period of evolution of the universe has also
been studied in some papers \cite{pert}.

As in LQC, the energy density and the Hubble parameter are
bounded, the future singularity arisen in the dark energy models
may also be avoided in this model.

The most adopted source for the matter in inflation scenario is a
scalar field with a suitable potential \cite{inf}. However,
obtaining an exact solution to the modified Friedmann equations in
LQC is not possible for a general potential. Inflation, in the
context of general relativity, may be driven by a slow roll scalar
field whose parameters are so fine tuned to provide enough number
of e-folding according to astrophysical data \cite{inf}. The slow
roll solution exists also in LQC, and by using numerical methods
it has been shown that the problem of fine tuning is alleviated in
this model \cite{inf1}.

To investigate the relationship between thermodynamics and
gravity,  many studies about thermodynamics properties of
cosmological event horizons and thermal nature of the enclosed
matter have been done in recent years \cite{therm}. In this
regard, validity of the second and generalized second laws (GSL)
of thermodynamics in different models of gravity has been subject
of many studies \cite{therm1}. Some of these works have been
inspired by the attempts to know the nature of matter with
negative pressure such as dark energy and inflaton causing the
acceleration or super-acceleration of the universe \cite{therm2}.
The scheme of the paper is a follows.

We consider a spatially flat homogeneous and isotropic
Friedmann–-Lema\^{i}tre–-Robertson–-Walker (FLRW) space-time and
focus on inflation era and specially on transition from
super-acceleration to acceleration epoch. After a brief
introduction to loop quantum cosmology and one of its exact
solutions, i.e. single fluid solution with constant equation of
state parameter (EoS), we present a slow rolling scalar field
solution, with power law potential, in the era of transition from
super-inflation to inflation phase. We show that this slow roll
solution is capable to describe the transition from
super-acceleration to acceleration expansion, provided that some
conditions which are compatible with Planck 2013 data are
satisfied. Afterward, we study the behavior of the entropy and
investigate the validity of the second and generalized second laws
of thermodynamics for our proposed solutions. In our study the
pre-bounce era is not considered.

We use natural units $\hbar=c=G=1$ throughout the paper.

\section{loop quantum cosmology, preliminaries, solutions}

We consider a spatially flat FLRW space-time
\begin{equation}\label{1}
ds^2=-dt^2+a^2(t)(dx^2+dy^2+dz^2),
\end{equation}
where $a(t)$ is the scale factor. By coupling the matter to
classical phase space of FLRW universe, and by taking into account
the constraints coming from loop quantum gravity, one obtains
\cite{lqc1}:
\begin{eqnarray}\label{2}
\dot{v}&=&{3v\over \lambda \gamma}\sin(\lambda b)\cos (\lambda
b)\nonumber \\
\mathcal{H}_{matt}&=&{3\over 4\gamma\lambda^2}v \sin^2(\lambda b).
\end{eqnarray}
$\mathcal{H}_{matt}$ is the matter Hamiltonian. $\gamma\approx
0.24$ \cite{ent} denotes the Barbero-Immirzi parameter and
$\lambda$ is the length gap $\lambda\approx 2.27$. $v$ is
proportional to the physical volume of a cubical cell, with unit
comoving volume : $v={a^3\over 2\pi\gamma}$, and $b$ is the
conjugate momentum. Dot denotes derivative with respect to the
time "$t$". The above equations can be rewritten as
\begin{eqnarray}\label{3}
H&=&{\sin(2\lambda b)\over 2\gamma\lambda} \nonumber \\
{\sin^2(\lambda b)\over \gamma^2 \lambda^2}&=&{8\pi\over 3}\rho,
\end{eqnarray}
where $H$ is the Hubble parameter and the matter density, $\rho$,
has an upper limit $\rho\leq \rho_c$ where $\rho_c={3\over 8\pi
\gamma^2\lambda^2}\approx 0.41$. (\ref{3}) results in the modified
Friedmann equation
\begin{equation}\label{4}
H^2={8\pi\over 3}\rho(1-{\rho\over \rho_c}).
\end{equation}
The continuity equation for the matter is
\begin{equation}\label{5}
\dot{\rho}+3H(P+\rho)=0,
\end{equation}
which by combining with (\ref{4}), gives the other Friedmann
equation
\begin{equation}\label{6}
\dot{H}=-4\pi(P+\rho)(1-{2\rho\over \rho_c}),
\end{equation}
where $P$ is the matter pressure. Although for ${\rho\ll \rho_c}$
we obtain the usual Friedmann equations, but in the Planck regime,
where $\rho\sim \rho_c$, the behavior of the system changes
drastically.

From (\ref{4}) and (\ref{6}), we can describe the expansion of a
universe filled with a perfect fluid satisfying $P+\rho>0$ as
follows: A bounce occurs at $\rho=\rho_c$, when $H=0$. After the
bounce $\rho$ decreases and in the interval ${\rho_c\over
2}<\rho<\rho_c$, the universe undergoes a super-inflation
evolution in the sense that $\dot{H}>0$. Note that the inflation
era is specified by a positive accelerated expansion:
$\ddot{a}>0$, which in terms of the Hubble parameter, is
equivalent to $\dot{H}+H^2>0$. Inflation during $\dot{H}>(<)0$ is
dubbed as super(normal)-inflation. Equipped with modified
Friedmann equations (i. e. eqs. (\ref{4}) and (\ref{6})), in order
to have a super-inflation phase there is no need to introduce a
phantom fluid satisfying $P+\rho<0$. During the super-inflation,
(\ref{5}) dictates that $\rho$ continues to decrease. When
$\rho={\rho_c\over 2}$, we have $\dot{H}=0$. Afterwards
$\rho<{\rho_c\over 2}$ and the universe undergoes a normal
inflation, i.e. $\ddot{a}>0$ but $\dot{H}<0$. The time at which
$\rho={\rho_c\over 2}$ is dubbed transition time. If $\lim_{t\to
\infty}\rho(t)={\rho_c\over 2}$, the super-inflation lasts
forever.

Combining (\ref{4}) and (\ref{6}) we get
\begin{equation}\label{ref1}
\dot{H}+H^2={4\pi\rho\over 3}\left[-3w-1+{2\rho\over
\rho_c}(2+3w)\right],
\end{equation}
where $w$ is the effective equation of state parameter defined by
$w={P\over \rho}$. Inflation occurs when
\begin{equation}\label{ref2}
3w+1<{2\rho\over \rho_c}(2+3w).
\end{equation}
In the context of general relativity, (\ref{ref2}) reduces to
$w<-{1\over 3}$, so to describe the inflation we were obliged to
introduce matter with negative pressure such as a slowly rolling
scalar field (inflaton). But in the framework of {\it {modified
Friedmann equations}} in loop quantum cosmology, even barotropic
fluids with non-negative constant EoS parameters such as radiation
whose densities satisfy (\ref{ref2}) not only can give rise to
inflation but also to super-inflation \cite{Sloan}. To get some
intuition about this point, consider a universe filled with a
perfect fluid whose the EoS parameter is a constant. $\rho$ is
obtained as
\begin{equation}\label{7}
\rho={\rho_c\over 1+6\pi \rho_c (1+w)^2 t^2}.
\end{equation}
In (\ref{7}) the time is adjusted such that the bounce happens at
$t=0$. The transition time, $t_0$ (defined as the time when the
super-inflation ($\dot{H}>0$) ends and normal inflation
($\dot{H}<0)$ begins), is given by
\begin{equation}\label{8}
t_0=\sqrt{1\over 6\pi\rho_c (w+1)^2}.
\end{equation}

A common candidate for the source of inflation in the literature
is a scalar field (inflaton) $\phi$. Choosing the potential as
\cite{new1}
\begin{equation}\label{ref3}
V(\phi)={(1-w)V_0e^{-\sqrt{24\pi(1+w)}\phi}\over \left(1+{V_0\over
2\rho_c}e^{-\sqrt{24\pi(1+w)}\phi}\right)^2},
\end{equation}
where $V_0$ is a constant, the scalar field mimics classically the
behavior of a perfect fluid with constant EoS parameter $w$. When
the scalar field is without potential we have $w=1$ (to find
discussions about inflation in this situation see \cite{Bojo}).
The model with $w=0$ (dust like behavior), and its cosmological
results, were studied in \cite{new1}.

\begin{figure}[H]
\centering\epsfig{file=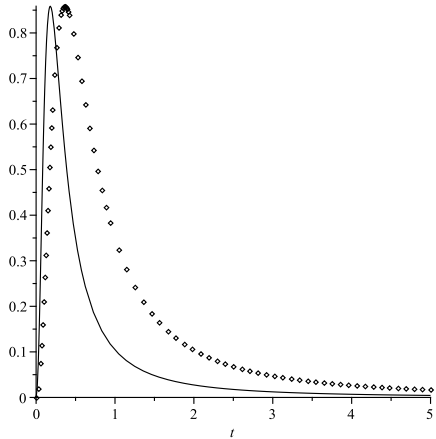,width=6cm,angle=0} \caption{$H^2$,
in terms of cosmic time for $w={1} (line)$, and $w=0$
(points)}\label{fig1}
\end{figure}

In fig. (\ref{fig1}), the Hubble parameter is depicted in terms of
the cosmic time for a universe filled with a perfect fluid whose
EoS is $w=1$, and also for $w=0$.

The bounce is assumed to occur at $t=0$ and we have set
$\rho_c\approx 0.41$, obtained by the Immirzi factor
$\gamma\approx 0.24$ derived by studying the black hole entropy in
loop Quantum Gravity in \cite{ent}. As it is illustrated in this
figure, after the bounce at $t=0$, we have a super-inflation phase
where $H$ is increasing. After $\dot{H}=0$, normal inflation phase
begins. The super-inflation lasts longer for smaller $w$.

Obtaining an exact analytical solution for (\ref{4}), and
(\ref{6}) is not generally possible. Indeed, (\ref{5}) can not be
easily integrated to give an analytic expression for $\rho(t)$
(note that only two of the equations (\ref{4}), (\ref{5}), and
(\ref{6}) are independent) but, as we have seen, choosing a
constant EoS parameter facilitates the analysis.

To generalize the discussion to no constant EoS parameters, in the
following we present a scalar field solution for the inflaton with
power law potential which crosses $\dot{H}=0$ line, during its
slow roll evolution. In general relativity, the slow roll model
can only describe the inflation phase but here, as we will show,
the super-inflation can occur also in the slow roll regime.

\subsection{ Slow roll scalar field solution crossing  $\dot{H}=0$ line }

In this part we assume that the inflaton is a slow-roll scalar
field with a power law potential. We seek for solutions of
Friedmann equations with differentiable Hubble parameter at
transition time which as before we denote by $t_0$. We present a
series which crosses $\dot{H}=0$ line, and show that by choosing
appropriate parameters it satisfies the Friedmann equations.

The energy density of the scalar field is
\begin{equation}\label{9}
\rho={1\over 2}\dot{\phi}^2+V(\phi)
\end{equation}
where $V(\phi)$ is the potential and the pressure is given by
\begin{equation}\label{10}
P={1\over 2}\dot{\phi}^2-V(\phi).
\end{equation}

In the neighborhood of $t=t_0$, we present the differentiable
Hubble parameter as \cite{ali}
\begin{equation}\label{11}
H(t)=h_0+h_1(t-t_0)^p+\mathcal{O}\left((t-t_0)^{p+1}\right),
\end{equation}
where $h_0$ is the value of the Hubble parameter and $p$ is the
order of the first non zero time derivative of $H$ at $t=t_0$. If
a negative $h_1={1\over p!}{d^pH\over dt^p}(t_0)$ and an even
integer number $p$ are found such that (\ref{11}) satisfies the
modified Friedmann equations, then this solution describes the
transition from super-inflation to inflation, because for $t\leq
t_0$ we have $\dot{H}\geq 0$ while for $t\geq t_0$ we have
$\dot{H}\leq 0$. The differentiability of the Hubble parameter at
transition time implies that the energy density and the pressure
are well defined (see (\ref{4} and (\ref{6})).

To justify our ansatz (\ref{11}), we show that it satisfies
(\ref{5}) and (\ref{6}) consistently (note that (\ref{4}) can be
derived from (\ref{5}) and (\ref{6})). The continuity equation for
the scalar field reads
\begin{equation}\label{12}
\ddot{\phi}+3H\dot{\phi}+V'(\phi)=0,
\end{equation}
where $V'(\phi)={dV(\phi)\over d\phi}$. Substituting (\ref{11})
into (\ref{12}), with the assumption $\ddot{\phi}\ll
3H\dot{\phi}$, yields
\begin{equation}\label{13}
3\left(h_0+h_1(t-t_0)^p+\mathcal{O}\left((t-t_0)^{p+1}\right)\right)\dot{\phi}+V'(\phi)=0,
\end{equation}
whose the solution for the power law potential
\begin{equation}\label{14}
V(\phi)=v_1\phi^n,
\end{equation}
where $v_1$ is a real number and $n$ is an integer, is given by
\begin{equation}\label{15}
\phi=\left({n(n-2)v_1(t-t_0)\Phi(-{{(t-t_0)^ph_1\over
h_0}},1,{1\over p})\over 3h_0 p}+c_1\right)^{1\over 2-n}.
\end{equation}
$c_1$ is determined by
\begin{equation}\label{16}
c_1=\phi^{2-n}(t_0),
\end{equation}
and $\Phi$ is the Lerchphi function
\begin{equation}\label{17}
\Phi(x,a,b)=\sum_{n=0}^\infty{x^n\over (b+n)^a}.
\end{equation}
For quadratic potential,
\begin{equation}\label{18}
V(\phi)={1\over 2}m^2\phi^2,
\end{equation}
we obtain
\begin{equation}\label{19}
\phi=\phi(t_0)\exp\left[{-m^2(t-t_0)\Phi\left(-{(t-t_0)^ph_1\over
h_0},1,{1\over p}\right)\over 3h_0p}\right].
\end{equation}
The slowly varying condition  $\ddot{\phi}\ll 3H\dot{\phi}$, used
to obtain the above solution, is satisfied when
\begin{equation}\label{20}
\left|{n(n-1)v_1\over 9h_0^2}\phi^{n-2}(t_0)\right|\ll  1,
\end{equation}
which for quadratic potential gives $m^2\ll h_0^2$, leading to
\begin{equation}\label{wmap1}
m^2\ll \rho_c.
\end{equation}
If we rewrite (\ref{20}) as $\left|{d^2V\over d\phi^2}\right|\ll
h_0^2$, in our slow roll regime, we deduce that the potential must
satisfy $\left|{d^2V\over d\phi^2}\right|\ll V$, implying that the
potential must be flat enough. Combining (\ref{20}) with
\begin{equation}\label{21}
v_1\phi^n(t_0)\simeq {\rho_c\over 2}
\end{equation}
(which means that in the slow roll approximation the main part of
the energy density is coming from the potential), yields
\begin{equation}\label{22}
\left|{n(n-1)\over 12\pi }\right|\ll \phi^2(t_0).
\end{equation}
So the value of the scalar field at transition time must be large
in this approximation.

By inserting (\ref{15}) into the Friedmann equation
\begin{equation}\label{23}
\dot{H}=-4\pi\dot{\phi}^2\left(1-{2\rho\over \rho_c}\right),
\end{equation}
we arrive at
\begin{eqnarray}\label{24}
&&\dot{H}=-{8\pi\over
27\rho_c}\left({n^2v_1^2\phi^{2(n-1)}(t_0)\over
h_0^3}\right)\left({n^3(n-1)v_1^3\phi^{3n-4}(t_0)\over
9h_0^2}+n^2v_1^2\phi^{2(n-1)}(t_0)\right)\nonumber \\
&&\times(t-t_0)+\mathcal{O}\left((t-t_0)^2\right)
\end{eqnarray}
So (\ref{11}) satisfies Friedmann equation provided that
\begin{eqnarray}\label{25}
h_1&=&-{4\pi\over 27\rho_c}\left({n^2v_1^2\phi^{2(n-1)}(t_0)\over
h_0^3}\right)\left({n^3(n-1)v_1^3\phi^{3n-4}(t_0)\over
9h_0^2}+n^2v_1^2\phi^{2(n-1)}(t_0)\right),\nonumber \\
p&=&2.
\end{eqnarray}
Applying the slow roll condition (\ref{20}) to (\ref{25}), we
obtain
\begin{equation}\label{26}
h_1\simeq-{4\pi\over 27
}{n^4v_1^4\left(\phi^4(t_0)\right)^{n-1}\over h_0^3\rho_c},
\end{equation}
which determines the transition rate. As $h_1<0$, our series
solution
\begin{eqnarray}\label{27}
H&\simeq&\sqrt{{2\pi\over 3}\rho_c}-\sqrt{2\over 27\pi}{n^4v_1^4\phi^{4(n-1)}(t_0)\over \rho_c^{5\over 2}}(t-t_0)^2+\mathcal{O}\left((t-t_0)^3\right)\nonumber \\
\rho&\simeq&{\rho_c\over 2}-{n^2v_1^2\phi^{2(n-1)}(t_0)\over
\sqrt{6\pi \rho_c}}(t-t_0)+\mathcal{O}\left((t-t_0)^2\right),
\end{eqnarray}
where $\phi(0)$ is determined by (\ref{21}), describes a
transition from super-inflation to inflation at $t=t_0$. Note that
the first term in the series solution of $H$ is the value of $H$
at the transition time, i.e. when $\rho={\rho_c\over 2}$. As a
summary, we conclude that crossing the super-inflation-inflation
divide line is, in principle, possible during a slow roll
evolution of a scalar field with power law potential, and the
approximate solution to Friedmann equations, in this region, is
given by (\ref{27}).

Compatibility of the slow roll scalar field model(with the power
law potential (\ref{18}))with observations was studied in
\cite{lqc1}. In \cite{lqc1} WMAP7 \cite{WMAP7} data were used to
obtain some estimation about the value of the scalar field and its
mass in the slow roll regime:
\begin{equation}\label{wmap2}
\phi(t_{*})=3.15m_{P},\,\,\,m=1.21\times10^{-6}m_{P}.
\end{equation}
$m_P$ is the Planck mass $m_P=\sqrt{\hbar c\over G}$. In the units
adopted in this paper $m_P=1$.  $t_{*}$ is some time in slow roll
era, after super-inflation where the scale length $\lambda={1\over
k}$ ($k=2\times 10^{-3}Mpc^{-1}$ is WMAP7 pivot scale) exited the
Hubble radius. We can follow the procedure used in \cite{lqc1}, to
update the results (\ref{wmap2}) via Planck 2013 data
\cite{planck2013}. In the scalar field model with square potential
(\ref{18}), the power spectrum of the scalar perturbations,
$\mathcal{P}_s$, and the spectral index, $n_s$, are specified as
\begin{eqnarray}\label{wmap3}
\mathcal{P}_s&=&{H^2\over \pi\epsilon}\nonumber \\
n_s&=&1-4\epsilon,
\end{eqnarray}
where
\begin{eqnarray}\label{wmap4}
\epsilon&=&{1\over 16\pi}\left({V'(\phi)\over
V(\phi)}\right)^2\nonumber \\
&=&{1\over 4\pi}{1\over \phi^2},
\end{eqnarray}
is one of the slow roll parameter.  (\ref{wmap3}) must be computed
at time $t_{*}$, where the scale length corresponding to the pivot
scale $k=0.05Mpc^{-1}$ exited the Hubble Horizon.  The Planck 2013
data imply (for $\%68$ CL or $1\sigma$ error)
\begin{eqnarray}\label{planck1}
\mathcal{P}_s&=&10^{-9}\times (09616\pm 0.16)\nonumber \\
n_s&=&0.9616\pm 0.0094.
\end{eqnarray}
(\ref{planck1}) together with (\ref{wmap3}) and (\ref{wmap4})
yield
\begin{eqnarray}\label{planck4}
&&4.71\times 10^{-11}<H^2(t_{*})<8.97\times 10^{-11} \nonumber
\\
&&6.66<\phi^2(t_{*})<10.98
\end{eqnarray}
and
\begin{equation}\label{planck5}
1.01\times 10^{-6}<m<1.73\times 10^{-6}.
\end{equation}

The estimated values for the inflaton mass in (\ref{planck5}) and
(\ref{wmap2})are in agrement with our result (\ref{wmap1}). By
noting that in the slow roll regime the main part of the energy is
the potential energy, and also by bearing in mind that the energy
decreases during the slow roll, we expect that the absolute value
of $\phi$ decreases after the transition,
$\left|\phi(t_{*})\right|<\left|\phi(t_0)\right|$. Therefore
(\ref{22}) is also in agreement with (\ref{wmap2}) and
(\ref{planck4}).

\section{Second and generalized second laws in loop quantum cosmology}

The study of thermodynamic laws of cosmological horizons may
provide us a mean to get more information about the solutions of
(modified) Friedmann equations which describe the inflation and
also the late time accelerated expansion of the universe. In this
part we try to investigate the domain of validity of the
generalized second law (GSL) (which asserts that the sum of the
entropies of the horizon and the matter it encloses is a non
decreasing function of cosmic time) in loop quantum cosmology for
the solutions studied in the previous section, and also study the
constraints that this law puts on these solutions.

The entropy of the apparent horizon $R_h$ in loop quantum gravity
is given by \cite{ent}
\begin{equation}\label{28}
S_h=\pi R_h^2+\pi \alpha \ln(\pi R_h^2)+ \beta,
\end{equation}
where $\alpha\sim \mathcal{O}(1)$ and $\beta$ are real constants
arisen from quantum corrections.

A natural choice for the cosmological horizon is the apparent
horizon $R_h={1\over H}$ . Note that $R_h$ does not exists for
$H=0$, but $R_h(H\neq 0)\in \mathbb{R}$. By adopting this choice
we obtain
\begin{equation}\label{29}
\dot{S_h}=-2\pi(1+\alpha H^2){\dot{H}\over H^3}.
\end{equation}
Near $H=0$, $S_h$ is very large and $\dot{S_h}\approx
-2\pi{\dot{H}\over H^3}$. So after the bounce, $S_h$ decreases
very fast, reducing $S_h$ to smaller finite values. In the super
accelerated expansion era, $H$ increases from $H=0$ at the bounce
to $H=\sqrt{2\pi \rho_c\over 3}$ at the transition time, hence
$S_h$ increases only for $-{3\over 2\pi\rho_c}<\alpha<0$, and when
$(1+\alpha H^2)<0$. After the transition, we have $\dot{H}<0$, and
$\dot{S_h}>0$ is valid for $(1+\alpha H^2)>0$.

To study the GSL,  we must also consider the contribution of
matter, which satisfies weak energy condition $\rho>0$,
$P+\rho\geq 0$, to entropy. We use the first law of thermodynamics
\begin{equation}\label{30}
dE=TdS_{in}-PdV,
\end{equation}
to obtain
\begin{equation}\label{31}
(P+\rho)dV+Vd\rho=TdS_{in}.
\end{equation}
$T>0$ is the temperature and $S_{in}$ is the entropy of the matter
inside the horizon. (\ref{32}) and (\ref{5}) result in
\begin{equation}\label{32}
\dot{S_{in}}=-4\pi {(P+\rho)\over TH^2}(1+{{\dot H}\over H^2}).
\end{equation}
During the inflation,  $1+{{\dot H}\over H^2}>0$, therefore
$\dot{S_{in}}<0$. After the inflation $S_{in}$ decreases.

By adding (\ref{32}) to (\ref{29}), we find that GSL is valid
whenever
\begin{equation}\label{33}
\dot{S}_{tot}=-2\pi(1+\alpha H^2){\dot{H}\over H^3}-4 \pi
{P+\rho\over TH^2}(1+{\dot{H}\over H^2})\geq 0.
\end{equation}

Provided that the weak energy condition is valid, immediately
after the bounce $S_{tot}$ quickly decreases. Afterward, the
energy density decreases and the system approaches to the
transition time. In the accelerated expansion era
($\dot{H}+H^2>0$), validity of GSL requires
\begin{equation}\label{34}
\dot{H}(1+\alpha H^2)<0,\,\,\, for \dot{H}>0,
\end{equation}
and
\begin{equation}\label{35}
\dot{H}(1+\alpha H^2)<0 ,\,\,\, for \dot{H}<0.
\end{equation}
By comparing (\ref{34}) and (\ref{35}), and assuming that $H$ is
continuous at transition time, the value of $\alpha$ is fixed as
\begin{equation}\label{10000}
\alpha=-{1\over H^2(t_0)}=-{3\over 2\pi \rho_c}\approx -1.16.
\end{equation}
But this is not the whole history: In the super accelerated
expansion epoch, $H$ increases until it reaches its maximum value
$H=\sqrt{{2\pi\rho_c\over 3}}$ at transition time.  But this
result is in contradiction with (\ref{34}) which implies
$H(t<t_0)>\sqrt{{2\pi\rho_c\over 3}}$. So we conclude that,
besides a time interval after the bounce, GSL  does not hold in
the whole of a connected region comprising super-inflation, and
ordinary inflation eras. Now let us have a closer look to the
behavior of the system in the neighborhood of transition time
$t_0$, when $\dot{S}_{tot}$ reduces to
\begin{equation}\label{36}
\dot{S}_{tot}(t=t_0)=-4 \pi {P+\rho\over TH^2}.
\end{equation}
The only way to save GSL in this region for ordinary matter (i.e.
matter satisfying weak energy condition), is to have
$P(t_0)+\rho(t_0)=0$. For the scalar field model,
$P+\rho=\dot{\phi}^2\geq 0$. Based on the solutions derived in the
previous section, the slow roll scalar field with power law
potential (\ref{14}) has the following behavior in neighborhood of
$\dot{H}=0$ :
\begin{equation}\label{37}
\dot{\phi}^2(t)={n^2v_1^2\phi^{2(n-1)}(t_0)\over 6\pi
\rho_c}-{2n^3(n-1)v_1^3\phi^{3n-4}(t_0)\over
\sqrt{216\pi^3}\rho_c^{3\over 2}}(t-t_0)+\mathcal{O}((t-t_0)^2).
\end{equation}
But in the slow roll approximation, $V(\phi)\approx {\rho_c\over
2}$, then $\phi(t_0)\neq 0$, and we have $\dot{S}_{tot}(t_0)<0$.
Therefore at $t_0$ and at least in the neighborhood of $t_0$ GSL
is violated.

At the end of inflation,$t_{end}$, $\dot{H}+H^2=0$, and
\begin{equation}\label{38}
\dot{S}_{tot}(t_{end})={2\pi\over H}(1+\alpha H^2).
\end{equation}
Hence GSL violation is ceased at this time provided that $1+\alpha
H^2>0$ which is always true for $\alpha>0$ (a positive value for
$\alpha$ was proposed by \cite{hod}).

To be more quantitative and to get more insights about the GSL let
us examine (\ref{33}) using some of the solutions of Friedmann
equations. To do this, we have to specify the matter temperature.
This temperature is assumed to be equal to the horizon temperature
which can be expressed in terms of surface gravity \cite{temp},
$\kappa=-H\left(1+{\dot{H}\over 2H^2}\right)$, as
\begin{equation}\label{39}
T=\left|{\kappa_H\over 2\pi}\right|= \left|{H\over
2\pi}\left(1+{\dot{H}\over 2H^2}\right)\right|.
\end{equation}
So GSL states
\begin{equation}\label{40}
\dot{S}_{tot}=2\pi{\dot{H}\over H^3}\left[-(1+\alpha
H^2)+\left({H^2+\dot{H}\over
2H^2+\dot{H}}\right)\left({2\rho_c\over
\rho_c-2\rho}\right)\right]\geq 0.
\end{equation}
Hence GSL holds whenever the above inequality is satisfied.

For a barotropic fluid with constant $w>0$, and by considering
$\rho<\rho_c$, one can show that (\ref{40}) reduces to
\begin{equation}\label{453}
\left(1-2x\right)\left[-1-\alpha_cx(1-x)+\left({2\over
1-2x}\right)\left({-3\Gamma(1-2x)+2(1-x)\over
4(1-x)-3\Gamma(1-2x)}\right)\right]\leq 0
\end{equation}
where $\Gamma=w+1$, and $x:={\rho\over \rho_c}$,
$\alpha_c:={8\pi\over 3}{\alpha \rho_c}$ are dimensionless
parameters. According to our previous claim, when $x\to 1$, the
GSL is obviously violated. The same occurs when $x=1/2$,
independently of the values of the parameters $\alpha$ and
$\Gamma>1$. for $x\to 0$ and $\alpha H^2\ll 1$, (\ref{40}) reduces
to $\pi {{\dot{H}}^2\over H^3}\geq 0 $ which is true. This is the
domain of classical theory of gravity far from the Planck era. In
this era GSL holds. The inflation ends at $x={3\Gamma-2\over
6\Gamma-2}$. In this region  $\alpha>0$ is a sufficient condition
for validity of GSL.

\begin{figure}[H]
\centering\epsfig{file=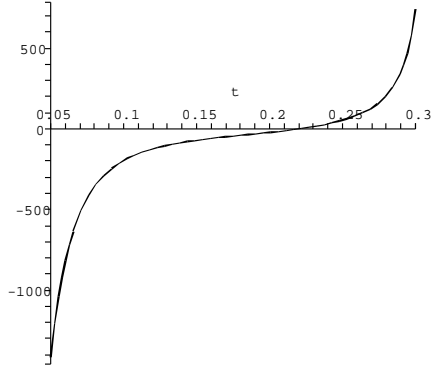,width=6cm,angle=0}
\caption{$\dot{S}_{tot}$ in terms of cosmic time for a single
fluid with $w=1$ and $\alpha=-1.16$}\label{fig2}
\end{figure}

In figure (\ref{fig2}), $\dot{S_{tot}}$ is depicted for a
barotropic fluid whose equation of state parameter is $w\approx 1$
(e.g. a massless scalar field).  GSL holds only for $t>0.22$ where
the inflation nearly ends. The transition time is $t_0=0.18$ (to
get more intuition, note these numbers are in terms of the
fundamental time unit, $\sqrt{\hbar G\over c^5}$)

\begin{figure}[H] \centering\epsfig{file=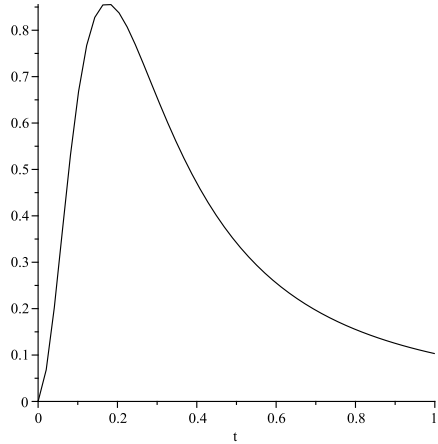,width=6cm,angle=0}
\caption{$H^2$ in terms of cosmic time for a scalar field model with
quadratic potential and extreme domination of kinetic energy at the
bounce}\label{fig3}
\end{figure}

\begin{figure}[H]
\centering\epsfig{file=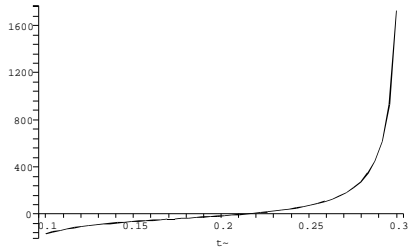,width=6cm,angle=0}
\caption{$\dot{S}$ in terms of cosmic time for a scalar field
model with quadratic potential and extreme domination of kinetic
energy at the bounce}\label{fig4}
\end{figure}

So far, to elucidate our results, we have used the slow roll
scalar field and barotropic fluids with constant EoS parameters.
To get a more complete insight about the behavior of entropy for a
nonconstant $w$, let us consider a scalar field with quadratic
potential, (\ref{18}), whose the kinetic energy is extremely
dominated at the bounce \cite{lqc1}. One of the characteristics of
this model is $f={\phi_B\over \phi_{max}}$, where
$\phi_{max}^2={2\rho_c \over m^2}$ and $\phi_B$ is the value of
the scalar field at the bounce. Therefore, $f^2={V(\phi)\over
\rho_{c}}$ determines the contribution of the potential in the
total energy density at the bounce. In \cite{lqc1}, it was shown
that for $f\geq 1.25\times 10^{-6}$ and $\phi_B>0$, this model is
consistent with WMAP data. We take $\phi_B=1.1$,
$\dot{\phi}_B=0.905$, $m=1.21\times 10^{-6}$ and $\alpha=-1.16$
which yield $f=1.4\times 10^{-6}$. As is depicted in
fig.(\ref{fig3})) after the bounce and until the transition time
$H$ is increasing, hence $\dot{H}+H^2>0$, thus this model is
enable to describe the onset of inflation.
In fig.(\ref{fig4})), $\dot{S}$  is depicted in terms of the
cosmic time . This figure shows that for $t< t\sim 0.2138$ GSL
fails.

\section{Conclusion}
We considered loop quantum cosmology, i.e. where the effective
quantum corrections coming from loop quantum gravity cannot be
neglected. We presented a slow roll scalar field solution with
power law potential in terms of Lerchphi function and showed that
this solution can describe the transition from the super-inflation
to the inflation phase, provided that the value of the scalar
field at transition time and the model parameters satisfy some
conditions (see (\ref{20}), (\ref{wmap1}) and (\ref{22})). It was
shown that these conditions are consistent with Planck 2013 data.
The thermodynamics second law and generalized second law (GSL) for
the apparent horizon and for some previously proposed solutions
were discussed. It was shown that if the matter respects the null
energy condition, immediately after the bounce the (total) entropy
decreases very fast from its huge values, but this decrease does
not continue for a long time and after awhile the entropy begins
to increase. It was shown that the GSL does not hold in the whole
of a connected region extended from super-inflation to inflation
era. Specially it was shown that for a scalar field this law is
violated at transition time. As illustrations, we discussed the
behavior of the entropy for an exactly solvable model (massless
scalar field) where GSL is violated even after the transition time
and holds (nearly) after the inflation ends. Afterwards using
numerical method we depicted time derivative of the total entropy
for scalar field model whose kinetic energy is extremely dominated
at the bounce, and the violation of GSL in a finite time interval
after the bounce was demonstrated.

So it seems that GSL does not hold near the bounce  where the
apparent horizon is very large, and  at least, depending on the
parameters of the model,  this violation may continue until the
end of inflation, as was illustrated through some examples. This
violation is related to effective quantum effects and GSL
continues to be true far from the Planck era. This may be related
to the fact that in our computation we have ignored the
contribution of the radiation energy density, generated by
particle creation from the horizon \cite{Modak}. Another
possibility to alleviate this problem may be the adoption of
another horizon such as the future event horizon \cite{last}.
Similar violation of  GSL was also reported in a super-accelerated
epoch in a phantom dominated universe or in modified theories of
gravity in the literature \cite{pav}.


\begin{thebibliography}{99}

\bibitem{lqc}A. Ashtekar, M. Bojowald, and J. Lewandowski, Adv.
Theor. Math. Phys. 7, 233 (2003); Abhay Ashtekar, Gen. Rel. Grav.
41, 741 (2009); A. Ashtekar, Nuovo Cim. 122B, 135 (2007),
arXiv:gr-qc/0702030.

\bibitem{sha}E. J. Copeland, D. J. Mulryne, N. J. Nunes, and M. Shaeri,
Phys. Rev. D 77, 023510 (2008).
\bibitem{pert} G. M. Hossain, Class. Quant. Grav. 22, 2511 (2005);
G. Calcagni and M. Cortes, Class. Quant. Grav. 24, 829 (2007); D.
J. Mulryne and N. J. Nunes, Phys. Rev. D 74, 083507 (2006); Y. S.
Piao and Y. Z. Zhang, Phys. Rev.  D 70, 063513 (2004).


\bibitem{inf} A. Linde,
Particle Physics and Inflationary Cosmology (Harwood, Chur,
Switzerland, 1990); A. Linde, Phys. Lett. B 129, 177 (1983).
\bibitem{inf1} A. Ashtekar, D. Sloan, Phys.Lett.B. 694, 108 (2010).
\bibitem{therm}T. Jacobson, Phys. Rev.
Lett. 75, 1260 (1995); R. G. Cai and S. P. Kim, J. High Energy
Phys. 02, 050 (2005) 050;  R. X. Miao, M. Li, and Y. G. Miao, JCAP
11, 033 (2011); Sh. F.  Wu, B. Wang,  G. H. Yang, and P. M. Zhang,
Class. Quant. Grav. 25, 235018 (2008); K. Bamba, M. Jamil, D.
Momeni, and R. Myrzakulov, arXiv:1202.6114v1 [physics.gen-ph]; V.
Faraoni , arXiv:1005.2327v1 [gr-qc]; A. Ashtekar and E.
Wilson-Ewing, Phys. Rev. D 78, 064047 (2008); K. Karami, A.
Abdolmaleki, N. Sahraei, and S. Ghaffari, J. High Energy Phys.
1108, 150 (2011); H. M. Sadjadi, Phys. Scripta 05, 055006 (2011);
H. M. Sadjadi and M. Jamil, Europhys. Lett 92, 69001 (2010); V.
Faraoni, A. F. Z. Moreno, and R. Nandra, arXiv:1202.0719v1
[gr-qc]; K. Karami, M.S. Khaledian, and N. Abdollahi,
arXiv:1201.4817v4 [physics.gen-ph];  H. M. Sadjadi , Europhys.
Lett. 92, 50014 (2010);  U. Debnath, Europhys. Lett. 94, 29001
(2011); V. Faraoni, Phys. Rev. D 80, 044013 (2009); M. Akbar, Int.
J. Theor. Phys. 48, 2665 (2009); A. Das, S. Chattopadhyay, and U.
Debnath, Found. Phys. 42, 266 ( 2011); Y. Zhang, Y. Gong, and Z.
H. Zhu, Int. J. Mod. Phys. 20, 1505 (2011); H. M. Sadjadi and M.
Honardoost, Phys. Lett. B 647, 231 (2007); Z. G. Liu and Y. S.
Piao, arXiv:1203.4901 [gr-qc]; Y. S. Piao and E. Zhou, Phys. Rev.
D 68, 083515 (2003).

\bibitem{therm1}
P. C.W. Davies, Class. Quant. Grav. 4, L255 (1987); 5, 1349
(1988);  D. Pavon, Class. Quant. Grav. 7, 487 (1990); R. Brustein,
Phys. Rev. Lett. 84, 2072 (2000); T. M. Davis, P. C. W. Davies,
and C. H. Lineweaver, Class. Quant. Grav. 20, 2753 (2003); K.
Bamba, R. Myrzakulov, S. Nojiri, and S. D. Odintsov,
arXiv:1202.4057v3 [physics.gen-ph]].

\bibitem{therm2} G. Izquierdo, D. Pavon, Phys. Lett. B 639, 1 (2006); M. D.
Pollock and T. P. Singh, Class. Quant. Grav. 6, 901 (1989); S.
Nojiri, S. D. Odintsov, Phys. Rev. D 72, 023003 (2005); H.
Farajollahi, A. Salehi, and F. Tayebi, Can. J. Phys. 89, 915
(2011); S. Nojiri and S. D. Odintsov,  Phys. Rev. D 70, 103522
(2004).
\bibitem{ent}K. A. Meissner, Class. Quant. Grav. 21, 5245 (2004),
A. Ghosh and P. Mitra, Phys. Rev. D 71, 027502 (2005).
\bibitem{Sloan}D. Sloan, "Loop Quantum Cosmology and the Early Universe", PhD Thesis,
The Pennsylvania State University (2010); X. Zhang, Y. Ling, JCAP
08, 012 (2007),  arXiv:0705.2656v2 [gr-qc].
\bibitem{new1}E. W. Ewing, JCAP 1303, 026 (2013), arXiv:1211.6269
[gr-qc].
\bibitem{Bojo}M. Bojowald, Living Rev. Rel. 8, 11 (2005); M.
Bojowald, F. Hinterleitner, Phys. Rev.  D 66 , 104003 (2002)
104003.
 \bibitem{lqc1}A. Ashtekar, and D. Sloan, Gen. Rel. Grav. 43, 3619
(2011).
\bibitem{ali}H. M. Sadjadi and M.
Alimohammadi, Phys. Rev. D 74, 043506 (2006).
\bibitem{WMAP7}E. Komatsu et al., Astrophys. J. Suppl. 192, 18 (2011), arXiv:1001.4538
[astro-ph.CO].
\bibitem{planck2013}P. A. R. Ade et al., Planck 2013 results. XVI, arXiv: 1303.5076
[astro-ph] (2013); P. A. R. Ade et al., Planck 2013 results. I,
arXiv: 1303.5062 [astro-ph) (2013).
\bibitem{hod} S. Hod, Class. Quant. Grav. 21, L97 (2004).
\bibitem{temp} R. G. Cai and
S. P. Kim, J. High Energy Phys.  0502, 050 (2005).
\bibitem{Modak} S. K. Modak and D. Singleton, arXiv:1205.3404v1
[gr-qc].
\bibitem{last}A. Ashtekar and E.
Wilson-Ewing, Phys. Rev. D 78, 064047 (2008).
\bibitem{pav}G. Izquierdo and D. Pavon, Phys. Lett. B 633, 420
(2006); H. Mohseni Sadjadi, Phys. Rev. D 73, 063525 (2006); H.
Mohseni Sadjadi, Europhys. Lett. 92, 50014 (2010).

\end{thebibliography}
\end{document}